\documentclass[PAPER, atlasdraft=false, UKenglish, cernpreprint, texmf, orcidlogo]{atlasdoc}
\usepackage{multirow}
\usepackage{threeparttable} %
\usepackage{graphicx}    %
\usepackage[subcaption]{atlaspackage}
\usepackage{atlasbiblatex}

\usepackage{atlasphysics}
\graphicspath{{logos/}{figures/}}

\usepackage{ANA-HIGP-2024-19-PAPER-defs}
\AtlasTitle{Search for the Higgs boson decay to a $Z$ boson and a photon in $pp$ collisions at $\sqrt{s}=13$~TeV and $13.6$~TeV with the ATLAS detector}
\AtlasAbstract{%
A search for the Higgs boson decay to a $Z$ boson and a photon in the $\ell\ell\gamma$ ($\ell = e, \mu$) final state is performed using $pp$ collisions at $\sqrt{s}=13.6$~TeV recorded with the ATLAS detector at the Large Hadron Collider during 2022--2024, corresponding to an integrated luminosity of 165~fb$^{-1}$.
The signal yield, normalised to the Standard Model prediction, is measured to be $\mu=0.9^{+0.7}_{-0.6}$, compatible with the expected value of $\mu=1.0\pm0.7$.
This corresponds to an observed (expected) signal significance of 1.4 (1.5) standard deviations under the background-only hypothesis.
This result is combined with that of a similar search performed with 140~fb$^{-1}$ of $\sqrt{s}=13$~TeV $pp$ collisions to provide the best expected sensitivity to date to this rare decay, namely an observed (expected) signal strength of $\mu = 1.3^{+0.6}_{-0.5}$ ($\mu = 1.0^{+0.6}_{-0.5}$), corresponding to an observed (expected) significance of 2.5 (1.9) standard deviations.
The measurement is consistent with the Standard Model expectation.}

\author{The ATLAS Collaboration}

\AtlasRefCode{HIGP-2024-19}

\PreprintIdNumber{CERN-EP-2025-155}

\AtlasJournalRef{\PLB 876 (2026) 140313}
\AtlasDOI{10.1016/j.physletb.2026.140313}

\AtlasCoverEgroupAnalysisTeam{atlas-higp-2024-19-analysis-team@cern.ch}

\hypersetup{pdftitle={ATLAS document},pdfauthor={The ATLAS Collaboration}}

\begin{document}

\maketitle

\section{Introduction}
\label{sec:intro}

% The next lines are included from the .//sections/Intro.tex input file
A new boson with properties consistent with those of the Standard Model (SM) Higgs boson ($H$) was independently observed in 2012 by the ATLAS and CMS Collaborations~\cite{HIGG-2012-27, CMS-HIG-12-028} at the Large Hadron Collider (LHC)~\cite{Evans:2008zzb}. Measurements of its couplings to SM particles, its spin, and its parity have, to date, shown no significant deviation from the SM predictions~\cite{HIGG-2021-23, CMS-HIG-22-001, HIGG-2013-17, CMS-HIG-14-018}. The combined ATLAS and CMS determination of its mass from LHC \ROne data yields $m_H = 125.09 \pm 0.24$~\GeV~\cite{HIGG-2014-14}, a value that is consistent with recent \RTwo measurements~\cite{HIGG-2022-20-Fixed, CMS-HIG-21-019-Fixed, CMS-HIG-19-004}.

Within the SM, the Higgs boson decay to a $Z$ boson and a photon ($H \to Z\gamma$) proceeds via loop-induced processes, leading to a predicted branching ratio of $\mathrm{BR}(H\to Z\gamma) = (1.54^{+0.10}_{-0.11})\times10^{-3}$ for $m_H = 125.09$~\GeV~\cite{deFlorian:2016spz}, similar to that for $H\to\gamma\gamma$, $\mathrm{BR}(H\to\gamma\gamma)=(2.27^{+0.07}_{-0.06}) \times 10^{-3}$~\cite{deFlorian:2016spz}.
The observation of this decay channel, not yet established, would not only complete the suite of established Higgs bosons into pairs of electroweak gauge bosons ($\gamma\gamma$, $ZZ^*$, $WW^*$) but also represent a sensitive probe for indirect signs of new physics. In fact, extensions of the SM, such as models with additional colourless charged scalars, fermions, or vector bosons, can alter this rate via contributions in the loops~\cite{carena, extensions, Chen:2013vi, Djouadi_1998, PhysRevD.100.075014, Archer_Smith_2021, He:2024bxi}, making the measurement of $\mathrm{BR}(H\to Z\gamma)$ and its ratio with $\mathrm{BR}(H\to\gamma\gamma)$ a stringent test of physics beyond the SM.
Similarly, scenarios in which the Higgs boson is composite or arises from alternative symmetry-breaking sectors may yield significantly different $H\to Z\gamma$ branching fractions~\cite{azatov, low, low2}.

The \Zlly final state ($\ell = e, \mu$), despite its reduced branching ratio, offers the best sensitivity to the $ H\to Z\gamma$ process, as it can be efficiently triggered on and clearly distinguished from background events produced in proton--proton ($pp$) collisions. Furthermore, it benefits from full kinematic reconstruction and excellent invariant mass resolution. Both the ATLAS and CMS Collaborations have performed searches for the $H \rightarrow Z(\rightarrow \ell\ell)\gamma$ decay using the \RTwo $pp$ collision dataset, corresponding to an integrated luminosity of about 140~\invfb at $\sqrt{s}=13$~\TeV for each experiment.
Compared to the background-only hypothesis, ATLAS observed an excess of events at $m_{H}=125.09~\mathrm{GeV}$ with a significance of $2.2\sigma$, while the expectation from the SM is $1.2\sigma$ with a measured signal strength, defined by the ratio of the signal yield to the SM prediction, $\mu = 2.0^{+1.0}_{-0.9}$~\cite{HIGG-2018-42}. Similarly, the CMS Collaboration reported an excess with a significance of $2.7\sigma$ ($1.2\sigma$ expected) with $\mu = 2.4\pm0.9$~\cite{CMS-HIG-19-014}.
The combination of these searches yielded the first evidence of \htoZg decays with a significance of $3.4\sigma$ ($1.6\sigma$ expected) with $\mu = 2.2\pm0.7$~\cite{HIGG-2022-22-fixed}. The combined result is consistent with the SM expectation within two standard deviations.

This letter presents a search for \htoZg decays in the \lly final state using $pp$ collision data collected with the ATLAS detector at $\sqrt{s}=13.6$~\TeV during 2022--2024 in Run~3, corresponding to an integrated luminosity of 165~\invfb.
Compared to the \RTwo analysis~\cite{HIGG-2018-42}, this study benefits from an increased Higgs boson production cross-section at higher centre-of-mass energy, and from a larger data and simulated background samples that improve the description of the expected background distributions in both the multivariate classifications and the background modelling studies by reducing statistical fluctuations and model uncertainty.
Furthermore, the event selection is optimised with relaxed transverse-momentum (\pt) thresholds for muons and photons.
The signal is extracted from 13 mutually exclusive event categories including, for the first time, a dedicated multi-lepton category designed to target Higgs boson production in association with a vector boson or top quarks.
The remaining 12 categories employ multivariate classifiers based on XGBoost~\cite{chen2016xgboost} to maximise the sensitivity, replacing the previous approach, mostly based on rectangular selections on simple kinematic variables.
A simultaneous fit to the reconstructed \Zg invariant mass distributions across all categories is performed to extract the overall \htoZg signal yield.
Finally, the result is combined with the \RTwo measurement to increase the sensitivity to this rare decay.
%

% End of text imported from the .//sections/Intro.tex input file

\FloatBarrier
\section{ATLAS detector}
\label{sec:detector}

% The next lines are included from the .//sections/Detector.tex input file
The ATLAS experiment~\cite{PERF-2007-01,GENR-2019-02} at the LHC is a multipurpose particle detector
with a forward--backward symmetric cylindrical geometry and near \(4\pi\) coverage in solid angle\footnote{ATLAS uses a right-handed coordinate system with its origin at the nominal interaction point (IP)
in the centre of the detector and the \(z\)-axis along the beam pipe.
The \(x\)-axis points from the IP to the centre of the LHC ring,
and the \(y\)-axis points upwards.
Polar coordinates \((r,\phi)\) are used in the transverse plane,
\(\phi\) being the azimuthal angle around the \(z\)-axis.
The pseudorapidity is defined in terms of the polar angle \(\theta\) as \(\eta = -\ln \tan(\theta/2)\) and is equal to the rapidity
$ y = \frac{1}{2} \ln \left( \frac{E + p_z}{E - p_z} \right) $ in the relativistic limit.
Angular distance is measured in units of \(\Delta R \equiv \sqrt{(\Delta y)^{2} + (\Delta\phi)^{2}}\).}.
It consists of an inner tracking detector (ID) surrounded by a thin superconducting solenoid
providing a \qty{2}{\tesla} axial magnetic field, electromagnetic (EM) and hadronic calorimeters, and a muon spectrometer (MS).
The inner tracking detector covers the pseudorapidity range \(|\eta| < 2.5\).
It consists of silicon pixels, silicon microstrips, and transition radiation tracking detectors.
Lead/liquid-argon (LAr) sampling calorimeters provide electromagnetic (EM) energy measurements
with high granularity within the region \(|\eta|< 3.2\).
A steel/scintillator-tile hadronic calorimeter covers the central pseudorapidity range (\(|\eta| < 1.7\)).
The endcap and forward regions are instrumented with LAr calorimeters
for EM and hadronic energy measurements up to \(|\eta| = 4.9\).
The muon spectrometer surrounds the calorimeters and is based on
three large superconducting air-core toroidal magnets with eight coils each.
The field integral of the toroids ranges between \num{2.0} and \qty{6.0}{\tesla\metre}
across most of the detector.
The muon spectrometer includes a system of precision tracking chambers up to \(|\eta| = 2.7\) and fast detectors for triggering up to \(|\eta| = 2.4\).
The luminosity is primarily measured by the LUCID-2 detector, which is located close to the beam pipe.
A two-level trigger system was used to select events~\cite{TRIG-2016-01,TRIG-2022-01}.
The first-level trigger is implemented in hardware and uses a subset of the detector information
to accept events at a rate close to \qty{100}{\kHz}.
This is followed by a software-based trigger that
reduced the accepted rate of complete events to \qty{3}{\kHz} on average, depending on the data-taking conditions.
A software suite~\cite{SOFT-2022-02} is used in data simulation, in the reconstruction
and analysis of real and simulated data, in detector operations, and in the trigger and data acquisition
systems of the experiment.

% End of text imported from the .//sections/Detector.tex input file

%
\FloatBarrier
\section{Data and simulation samples}
\label{sec:simu_sample}

% The next lines are included from the .//sections/Simulation_Samples.tex input file

The analysis presented in this letter uses $pp$ collision data at $\sqrt{s}=13.6$~\TeV collected by the ATLAS experiment from 2022 to 2024 during LHC Run~3.
Events were collected using unprescaled single- and dilepton triggers~\cite{TRIG-2018-05, TRIG-2018-01} with a variety of \pt thresholds.
The lowest-threshold single-electron and single-muon triggers required $\pt > 26$~\GeV and $\pt > 24$~\GeV, respectively. The dielectron trigger required two electrons with $\pt > 17$~\GeV each. The dimuon triggers employed a symmetric 14~\GeV--14~\GeV configuration in 2022--2024 and an asymmetric 22~\GeV--8~\GeV configuration in 2023--2024.
To improve the efficiency at high instantaneous luminosity, these low-threshold triggers were supplemented by higher-\pt triggers with looser identification or isolation requirements.
The trigger selections yield an efficiency of 96\% for the $ee\gamma$ final state and 93\% for the $\mu\mu\gamma$ final state for events passing the offline selection requirements that will be described in Section~\ref{sec:event_selection}. After data quality requirements, the total integrated luminosity amounts to \lumithirteensixtev~\invfb.
The average number of inelastic $pp$ interactions per bunch crossing (\pileup) increased from 42 in 2022 to 58 in 2024, with the peak instantaneous luminosity reaching \peakLumi.

The optimisation of the analysis strategy and the modelling of the relevant physics processes rely on simulation Monte Carlo (MC) samples that represent both the Higgs boson signal and the dominant background processes.
These samples, unless explicitly stated otherwise, undergo the full simulation of the ATLAS detector response using the $\GEANT$ framework~\cite{Geant4},
as implemented in the ATLAS simulation infrastructure~\cite{SOFT-2010-01}.

The Higgs boson mass is set to $\mh = 125$~\GeV for all simulated samples, with a corresponding total decay width of $\Gamma_{H} = 4.1$~\MeV, as recommended in Ref.~\cite{deFlorian:2016spz}.
The simulated samples are normalised to the SM production cross-sections, evaluated at a Higgs boson mass of $\mh = 125.09$~\GeV{}.
These include production via gluon--gluon fusion (\ggF)~\cite{deFlorian:2016spz,Anastasiou:2015ema,Anastasiou:2016cez,Dulat:2018rbf,Harlander:2009mq,Harlander:2009bw,Harlander:2009my,Pak:2009dg,Actis:2008ug,Actis:2008ts-Fixed,Anastasiou:2008tj,Aglietti:2004nj,Bonetti:2018ukf},
vector boson fusion (\VBF)~\cite{deFlorian:2016spz,Ciccolini:2007jr-Fixed,Ciccolini:2007ec,Bolzoni:2010xr},
associated production with a vector boson (\VH, where $V = W, Z$)~\cite{deFlorian:2016spz,Ciccolini:2003jy,Brein:2003wg,Brein:2011vx,Altenkamp:2012sx,Denner:2014cla,Brein:2012ne-Fixed,Harlander:2014wda,Harlander:2018yio},
associated production with a top-quark pair (\ttH)~\cite{Beenakker:2002nc,Dawson:2003zu,Yu:2014cka,Frixione:2015zaa,deFlorian:2016spz},
and associated production with a bottom-quark pair (\bbH)~\cite{wiesemann2015higgs,dittmaier2004higgs,dawson2004exclusive}. Other Higgs boson production mechanisms were not considered, as their contributions to the total Higgs boson production cross-section are at the level of $0.1\%$ or less.
The \htoZg branching ratio and its uncertainty are also taken from Ref.~\cite{deFlorian:2016spz}.
The production of the Higgs boson was simulated using the \POWHEGBOX~v2 MC event generator~\cite{Nason:2009ai,Alioli:2010xd,Nason:2004rx,Frixione:2007vw,Hartanto:2015uka} and the PDF4LHC21 parton distribution functions (PDF) set ~\cite{Ball:2022oua}, following the setup summarised in Table~\ref{tab:samples}.
The \htoZg decay, as well as parton shower, hadronisation, and the modelling of the underlying event, were performed using \PYTHIAV{8}~\cite{Bierlich:2022pfr-Fixed}.
Contributions from $H\rightarrow\mu\mu$ decays, where the reconstructed photon originates from QED final-state radiation (FSR), were evaluated using samples produced with a similar setup and are considered as a potential background in this analysis.
The impact of the interference between the signal and other Higgs boson decays with the same final-state signature (e.g. $H\rightarrow \gamma\gamma^{\star}\rightarrow\ell\ell\gamma$) is expected to be negligible in the SM~\cite{llyBR2013}.
Additional samples of the \HZy signals were generated using \HERWIGV{7}~\cite{bellm2020herwig} for decay and parton shower simulation and are used to evaluate uncertainties associated with the parton shower modelling.

In this analysis, the dominant backgrounds arise from non-resonant production of a $Z$ boson in association with a photon (\Zg) or jets, where a jet is misidentified as a photon (\zjets).
Additional background contributions stem from diboson ($VV$) production, where $V$ denotes either a $W$ or a $Z$ boson. These additional background contributions are relevant only for the event category containing additional leptons.

A large $Z\gamma$ background sample was generated at next-to-leading-order (NLO) accuracy in QCD using the \SHERPAV{2.2.14} generator~\cite{Bothmann:2019yzt}, with one to three additional partons in the final state at LO. The simulation employed the \NNPDF[3.0]~next-to-next-to-leading-order (NNLO) PDF set~\cite{Ball:2014uwa}, and a fast simulation of the calorimeter response was applied~\cite{SIMU-2018-04}.

The EW production of a $Z$ boson in association with a photon and two jets (\Zyjj) was simulated at LO using \MADGRAPHV{3.5.5}~\cite{Alwall:2014hca} with the \NNPDF[2.3LO] PDF set~\cite{ball2013parton}, where both jets originate from partons emitted at EW vertices.
QCD-induced diagrams were explicitly excluded. No additional partons are from QCD interactions in the final state, ensuring orthogonality with the $Z\gamma$ sample generated with \SHERPA. The hadronisation, parton shower, and the modelling of the underlying event were simulated using \PYTHIAV{8} with the A14 set of tuned parameters~\cite{ATL-PHYS-PUB-2014-021}.
The background from \zjets was estimated from data using a control region of events in which the photon candidates fail the nominal criteria and pass looser identification or isolation requirements.

Diboson backgrounds resulting in three- and four-lepton final states were also modelled using \SHERPAV{2.2.14}, employing the \NNPDF[3.0NNLO] PDF set. The sample was generated at NLO accuracy in QCD with up to three additional partons at LO accuracy.

The effect of \pileup was modeled by overlaying simulated hard-scattering events with inelastic $pp$ interactions from a mix of \EPOS[2.0.1.4]\cite{Werner:2005jf} and \PYTHIAV{8}. \EPOS (\PYTHIA) events were generated with the LHC~\cite{Pierog:2013ria} (A3~\cite{ATL-PHYS-PUB-2016-017}) tune and \NNPDF[2.3LO] PDF set. \PYTHIA simulated pileup events containing high-\pt jets, prompt photons, or leptons from $b$-hadron decays, while \EPOS accounted for the remaining minimum--bias contribution.
The individual samples were first reweighted to ensure a smooth transition across jet \pt, and the combined sample was reweighted to match the distribution of the number of interactions per bunch crossing observed in data.

To improve the agreement between simulation and data, correction factors were applied to account for differences in reconstruction and identification efficiencies. These include corrections to the trigger, reconstruction, identification, and isolation efficiencies for electrons and muons, as well as the identification and selection of photons and jets. Additional corrections were applied to the energy and momentum scale and resolution of reconstructed objects to reflect the detector performance.

\begin{DIFnomarkup}
\begin{table}[htbp]
\caption{Higgs boson signal MC samples produced with \POWHEGBOX~V2 along with the techniques used to generate the events and their precision in $\alpha_{\mathrm{s}}$ for the event generation (gen.). The version of \PYTHIAV{8.310}, A14 tune, and the PDF4LHC21 set, which are used for modelling the Higgs boson decay, parton shower, hadronisation, and the underlying event, are listed. The precision of the total cross-section used in the sample normalisation is also reported.}
\label{tab:samples}
\centering
\footnotesize
\begin{threeparttable}
\begin{tabular}{llll}
\hline \hline
Process     & Technique & QCD (gen.)  & Normalisation \\
\hline
\ggF        &
\POWHEG~\cite{Nason:2004rx,Frixione:2007vw} &NNLO\tnote{*}  &  N$^3$LO (QCD), NLO (EW)~\cite{Aglietti:2004nj,Actis:2008ug,Actis:2008ts-Fixed,Harlander:2009mq,Harlander:2009bw,Harlander:2009my,Pak:2009dg,Anastasiou:2008tj,Anastasiou:2015ema,Anastasiou:2016cez,Dulat:2018rbf,Bonetti:2018ukf} \\
\VBF        & \POWHEG & NLO & NNLO (QCD), NLO (EW)~\cite{Ciccolini:2007jr-Fixed,Ciccolini:2007ec,Bolzoni:2010xr} \\
\qqZH       & \POWHEG \& MiNLO~\cite{Luisoni:2013kna} & NLO & NNLO (QCD), NLO (EW)~\cite{Harlander:2018yio,Brein:2012ne-Fixed} \\
\ggZH       & \POWHEG & LO & NLO (EW)~\cite{Harlander:2018yio,altenkamp2013gluon,harlander2014soft} \\
\WH         & \POWHEG \& MiNLO~\cite{Luisoni:2013kna} & NLO & NNLO (QCD), NLO (EW)~\cite{Harlander:2018yio,Brein:2012ne-Fixed} \\
\ttH        & \POWHEG & NLO & NLO (QCD+EW)~\cite{Beenakker:2002nc,Dawson:2003zu, Yu:2014cka,Frixione:2015zaa,hartanto2015higgs} \\
\bbH        & \POWHEG (4FS) & NLO & NNLO (QCD), NLO (EW)~\cite{dittmaier2004higgs,dawson2004exclusive,wiesemann2015higgs,jager2016higgs} \\
\hline \hline
\end{tabular}
\begin{tablenotes}
\footnotesize
\item[*]
NNLO accuracy achieved only for inclusive \ggF observables.
\end{tablenotes}
\end{threeparttable}
\end{table}
\end{DIFnomarkup}
%
%
%
%
%
%
%
%
%
%
%
%
%
%
%
%
%
%
%
%
%
%
%
%
%
%
%
%
%
%
%
%
%
%
%
%
%
%
%
%
%
%

% End of text imported from the .//sections/Simulation_Samples.tex input file

%
\FloatBarrier
\section{Event selection and reconstruction}
\label{sec:event_selection}

% The next lines are included from the .//sections/Event_Selection.tex input file
Events passing the trigger selection described in Section~\ref{sec:simu_sample} are required to include at least one photon and two same-flavour, opposite-charge leptons, all associated with the primary vertex, defined as the inner-detector track vertex with the highest $\sum{\pt^2}$ of tracks with $\pt > 500$~\MeV~\cite{ATL-PHYS-PUB-2015-026}.

Muon candidates are reconstructed by combining the tracks in the ID and MS~\cite{MUON-2018-03, MUON-2022-01}.
They need to satisfy \medium identification criteria, lie within $|\eta| < 2.5$, and have $\pt > 5$~\GeV.
Electrons are reconstructed from topological clusters of EM calorimeter cells matched to ID tracks~\cite{EGAM-2018-01, EGAM-2021-01, EGAM-2021-02}, incorporating nearby electromagnetic energy deposits associated with bremsstrahlung photons.
They are required to pass a \loose identification selection based on a likelihood discriminant
using calorimeter shower shapes and track parameters, lie within $|\eta|<2.47$ (excluding the transition region between the barrel and endcap EM calorimeters $1.37<|\eta|<1.52$), and have $\pt > 10$~\GeV.
To ensure leptons originate from the primary vertex and to suppress heavy-flavour backgrounds, all leptons are required to satisfy $\zzsint < 0.5$~mm,
where $\Delta\zzero$ is the longitudinal impact parameter relative to the primary vertex and $\theta$ is the track polar angle.
Moreover, the transverse-impact-parameter significance $|d_{0}|/\sigma_{d_{0}}$ needs to be $<3$ for muons and $<5$ for electrons,
with \dzero measured relative to the beam line and $\sigma_{\dzero}$ its fit uncertainty.
Unconverted and converted photon candidates are built from topological clusters of EM calorimeter cells, possibly matched to tracks identified as candidates from $\gamma\to ee$ conversions ~\cite{EGAM-2018-01, EGAM-2021-01, EGAM-2021-02}. They have to satisfy \tight identification criteria based on calorimeter shower shape
variables, $|\eta|<2.37$ (excluding $1.37<|\eta|<1.52$), and have $\pt > 10$~\GeV (relaxed from 15~\GeV in Run~2~\cite{HIGG-2018-42}).
To increase the rejection of background from non-prompt and hadronic production, the lepton and photon candidates are required to pass isolation criteria, based on the energy deposits in the calorimeter and the total \pt of charged-particle tracks from the primary vertex, measured in cones surrounding the direction of the particle candidates.

Jets are reconstructed from particle flow objects~\cite{PERF-2015-09} clustered via the anti-$k_{t}$ algorithm ($R=0.4$)~\cite{Cacciari:2008gp, Fastjet, JETM-2018-05}. They are required to satisfy $\pt >25$~\GeV and $|y|<4.4$.
To suppress jets originating from pile-up, a central~\cite{PERF-2014-03} or forward jet-vertex-tagger~\cite{ATL-PHYS-PUB-2019-026}, requiring jets to come from the primary vertex, is applied.

Overlap removal discards the lower-\pt electron when two electron candidates share the same track or satisfy $|\Delta\eta|<0.075$ and $|\Delta\phi|<0.125$. Electrons within $\Delta R<0.02$ of muons are also removed. To suppress bremsstrahlung photons emitted by leptons, photons within $\Delta R<0.3$ of leptons are discarded. Finally, jets within $\Delta R<0.2$ of leptons or photons are removed.

$Z$ boson candidates are reconstructed from opposite-charge and same-flavour lepton pairs.
In the muon channel, the highest \pt collinear FSR photon ($\Delta R<0.15$) is added to the corresponding muon to improve the di-muon mass resolution.
In the electron channel, the reconstruction and calibration already account for radiative energy losses~\cite{EGAM-2021-01}.
A kinematic fit~\cite{HIGG-2018-42} then corrects the lepton four-momenta to constrain the dilepton mass to the known $Z$ boson mass, accounting for its finite natural width.
This procedure improves the $\ell\ell\gamma$ mass resolution of the signal MC samples by 17\% for electrons and 11\% for muons, including the effect of FSR.
Candidates must satisfy $|m_{\ell\ell}-m_{Z}|<10$~\GeV, where $m_{Z}=91.2$~\GeV~\cite{Zyla:2020zbs-fixed}. If multiple $Z$ boson candidates are found, the dilepton pair with $m_{\ell\ell}$ closest to $m_{Z}$ is selected, resulting in a 98\% efficiency for correctly matching to the true $Z$ boson in the ggF signal sample.
The leptons associated with the $Z$ boson candidate are additionally required to be geometrically matched to the corresponding trigger-level leptons that fired the event. They are required to satisfy an offline \pt threshold set 1--2~\GeV above the nominal trigger requirement to ensure that the trigger is maximally efficient. %

The Higgs boson candidate is reconstructed by combining the selected $Z$ boson and the highest--\pt photon.
The invariant mass of the $\ell\ell\gamma$ system (\mzg) must lie within 110--160~\GeV to suppress contributions from on-shell $Z$ boson events.
In addition, the photon \pt is required to be larger than 0.09 times \mzg. This requirement further reduces the background, while not sculpting the \mzg spectrum near $m_{H}$ as would the use of an absolute minimum $\pt^{\gamma}$ requirement with similar background rejection.
This criterion is relaxed from the \RTwo value of 0.12 times \mzg thanks to dedicated efficiency corrections and energy calibrations for photons with lower \pt.
The overall reconstruction and selection efficiency (including detector acceptance) for the SM \HZlly events ranges from 20\% to 26\%, depending on the production mode.
%

% End of text imported from the .//sections/Event_Selection.tex input file

%
\FloatBarrier
\section{Event categorisation}
\label{sec:event_categorization}

\begin{figure}[htbp]
\centering
\includegraphics[width=0.7\textwidth]{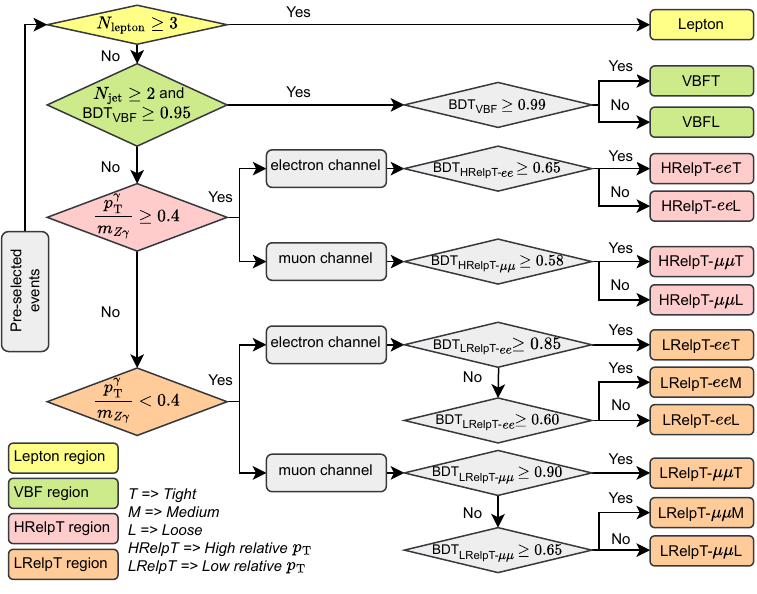}
\caption{Schematic of the event categorisation based on kinematic selections and multivariate algorithms. Events are classified into four regions: Lepton (yellow), VBF (green), high relative photon \pt (HRelpT, pink), and low relative photon \pt (LRelpT, orange).
The Lepton region is defined based on the number of lepton candidates, while the VBF region is split into loose and tight categories using a BDT output. The HRelpT and LRelpT regions are further divided into various categories based on the lepton flavour and dedicated BDT classifiers. In total, 13 mutually exclusive categories are defined.}
\label{fig:analysis_strategy}
\end{figure}

% The next lines are included from the .//sections/Categorization.tex input file
To maximise the sensitivity to an \HZy signal, events are classified into 13 exclusive event categories.
Events are first categorised into four primary regions based on lepton and jet multiplicities as well as the photon \pt relative to $m_{Z\gamma}$: Lepton, VBF, High relative photon \pt, and Low relative photon \pt.
The VBF and both the High and Low relative photon \pt regions are further subdivided into mutually exclusive subcategories with different signal purity and roughly similar significance according to the output of an XGBoost-based~\cite{chen2016xgboost} boosted decision trees (BDT) and to lepton flavour.
To avoid sculpting of the $m_{Z\gamma}$ distribution, the BDT inputs are selected to have minimal correlation with $m_{Z\gamma}$, and the classifiers are trained within narrow mass windows around the expected \mH: 120--130~\GeV for the VBF and High relative photon \pt regions, and 123--127~\GeV for the Low relative photon \pt region.
The number and boundaries of the categories are optimised to maximise the combined expected significance. Each category is required to contain at least two expected background events in the signal region ($120 < m_{Z\gamma} < 130$~\GeV).
This requirement translates into a minimum number of expected background events in the fit mass region of each category that is large enough to ensure sufficient statistical precision for the background estimate.
The event classification scheme is illustrated in Figure~\ref{fig:analysis_strategy}, and more details are provided in the following paragraphs.

Events with at least three leptons ($N_\text{lepton}\geq 3$) are assigned to the Lepton category. This category is enriched in \VH and \ttH production modes, accounting for 56\% and 35\% of the signal yield in this category, respectively. The dominant backgrounds arise from \zjets, multi-lepton diboson, and non-resonant \Zg processes.

Events failing the Lepton region and containing at least two jets ($N_\text{jet}\geq 2$) are selected for the VBF region. If more than two jets are present in an event, the two highest-\pT jets are considered. A dedicated BDT was trained using 25 kinematic variables to separate the \VBF signal from the non-resonant $Z\gamma$, and the EW $Z\gamma jj$, as well as to reduce the contamination from \ggF. The variables are summarised in Table~\ref{tab:BDT_inputs}.
Events in the VBF region are classified into tight (VBFT) and loose (VBFL) categories based on the BDT score, with the VBFT category exhibiting higher significance. The VBF fraction of the total signal yield is 89\% and 64\%, respectively, in the VBFT and VBFL categories.

\begin{table}[htbp]
\centering
\caption{Summary of input variables used in the BDT training for the VBF, \HRelpTee, \HRelpTmumu, \LRelpTee, and \LRelpTmumu analysis regions. The same set of input variables is employed in the four relative photon \pT regions.}
\resizebox{\textwidth}{!}{
\renewcommand{\arraystretch}{1.2}
\begin{tabular}{llcc}
\toprule
Input variable & Description & VBF & relative photon \pT \\
\midrule
$N_{\mathrm{jet}}$ & Number of jets & \checkmark & \checkmark \\
$p_\text{T}^{j_1}$, $\eta_{j_1}$ & \pT, and $\eta$ of the leading jet ($j_1$) & \checkmark & \checkmark \\
$m_{j_1}$ & Mass of the leading jet & \checkmark & \\
$\phi_{j_1}$ & Azimuthal angle of the leading jet & & \checkmark \\
$m_{j_2}$, $p_\text{T}^{j_2}$, $\eta_{j_2}$ & Mass, \pT, and $\eta$ of the subleading jet ($j_2$) & \checkmark &  \\
$m_{j_1j_2}$, $p_\text{T}^{j_1j_2}$, $\Delta\eta_{j_1j_2}$ & Dijet invariant mass, \pT, and $\eta$ separation & \checkmark &  \\

$\eta_{\gamma}$ & Pseudorapidity of the photon & \checkmark & \checkmark \\
$p_\text{T}^{\gamma}$ & Transverse momentum of the photon & \checkmark & \\
$p_\text{T}^{\ell\ell}$ & Transverse momentum of the dilepton system & \checkmark &  \\
$\eta_{\ell\ell}$, $\eta_{\ell\ell\gamma}$ & Pseudorapidity of the dilepton and $\ell\ell\gamma$ systems & \checkmark & \checkmark \\
$p_\text{T}^{\ell\ell\gamma}$ & Transverse momentum of the $\ell\ell\gamma$ system & \checkmark &  \\
$p_\text{T}^{\ell\ell\gamma}/m_{\ell\ell\gamma}$, $p_\text{T}^{\gamma}/m_{\ell\ell\gamma}$ & Relative \pT of the $\ell\ell\gamma$ and the photon &  & \checkmark \\
$p_\text{T}^{\ell\ell}/m_{\ell\ell\gamma}$ & Relative \pT of the dilepton system &  & \checkmark \\
\ptt & \begin{tabular}[t]{@{}l@{}} Component of $\vec{p}_\text{T}^{\ell\ell\gamma}$ perpendicular to the difference between \\ $\vec{p}_\text{T}^{\,\ell\ell}$ and $\vec{p}_\text{T}^{\,\gamma}$ ($\ptt = |\vec{p}_\mathrm{T}^{\ell\ell\gamma} \times \hat{t} |$, with $\hat{t} \propto \vec{p}_\text{T}^{\,\ell\ell} - \vec{p}_\text{T}^{\,\gamma}$)~\cite{Ackerstaff:1997rc,VESTERINEN2009432} \end{tabular}
& \checkmark & \checkmark \\

$\Delta\phi_{\ell\ell,\gamma}$ & Azimuthal separation between the $\ell\ell$ system and the photon & \checkmark & \checkmark \\
$\Delta\phi_{\ell\ell\gamma,j_1}$ & Azimuthal separation between the $\ell\ell\gamma$ system and the leading jet & \checkmark & \checkmark \\
$\Delta\phi_{\ell\ell\gamma,j_1j_2}$ & Azimuthal separation between the $\ell\ell\gamma$ system and the dijet system & \checkmark &  \\
$\Delta\eta_{\ell\ell,\gamma}$ & Pseudorapidity separation of the dilepton and the photon & \checkmark & \checkmark \\
$\Delta R^{\min}_{\gamma \text{ or } \ell\ell, j}$ & Minimum $\Delta R$ to $j_1/j_2$ from the photon or the dilepton system& \checkmark & \checkmark \\
$\cos\theta^*(\ell^+)$ & Cosine of the polar angle of the $\ell^+$ in the $\ell\ell$ rest frame & \checkmark & \checkmark \\
$\cos\theta(\ell\ell)$ in $\ell\ell\gamma$ & Cosine of the polar angle of the $\ell\ell$ in the $\ell\ell\gamma$ rest frame & \checkmark & \checkmark \\
$\eta^{\text{Zeppenfeld}}$ &
\begin{tabular}[t]{@{}l@{}}Pseudorapidity difference between the $\ell\ell\gamma$ system and \\ the dijet system, defined as $|\eta_{\ell\ell\gamma} - (\eta_{j_1} + \eta_{j_2})/2|$~\cite{Zeppenfeld}\end{tabular}
& \checkmark &  \\
\bottomrule
\end{tabular}
}
\label{tab:BDT_inputs}
\end{table}

Events not selected for the VBF region and satisfying  $\pt^\gamma / m_{Z\gamma} \geq 0.4$ enter the High relative photon \pt region, dominated by \ggF production. Separate BDTs are trained for the $ee$ and $\mu\mu$ channels using 17 variables, as shown in Table~\ref{tab:BDT_inputs}.
BDTs are trained using the simulated \ggF and \VBF signal events and the non-resonant $Z\gamma$ background events.
For each lepton flavour, events are grouped into tight and loose categories using optimised BDT thresholds, yielding four categories (\HRelpTeeT, \HRelpTeeL, \HRelpTmumuT, \HRelpTmumuL) with \ggF signal fractions above 70\%.

Remaining events that are not selected in the VBF region and with $\pt^\gamma / m_{Z\gamma} < 0.4$, are placed into the Low relative photon \pt region, where the signal yield is also dominated by \ggF with a fraction above 90\%. BDTs trained separately for $ee$ and $\mu\mu$ channels use the same input variables as in the High relative photon \pt region.
The training samples consist of the simulated \ggF signal events, the non-resonant $Z\gamma$ background, and the data-driven $Z$+jets background events.
Background components are constructed and normalised as detailed in Section~\ref{sec:sig_bkg_modeling}.
Each channel is split into tight, medium, and loose categories, according to their BDT score, forming six in total: \LRelpTeeT, \LRelpTeeM, \LRelpTeeL, \LRelpTmumuT, \LRelpTmumuM, and \LRelpTmumuL.

Table~\ref{tab:SBtable} provides an overview of the expected signal and background composition in each analysis category, as determined from a signal-plus-background fit to the \Zg invariant mass distribution of the Asimov dataset (Section \ref{sec:result}) to illustrate the sensitivity.
The expected signal ($S^\text{exp}_{68}$), background ($B^\text{exp}_{68}$), and observed data ($N_{68}$) yields are quoted in an $m_{Z\gamma}$ window around the peak, with a width ($w_{68}$) corresponding to the interval expected to contain 68\% of the signal.
The resulting expected sensitivity, $S^\text{exp}_{68}/\sqrt{S^\text{exp}_{68} + B^\text{exp}_{68}}$,
varies across categories, reaching a maximum of 0.91 in the VBFT category. The combined sensitivity, obtained by summing the contributions from all categories in quadrature, is 1.81. The inclusive sensitivity is used solely for category optimisation and is not intended to represent the final sensitivity derived from the full statistical analysis.

\begin{DIFnomarkup}
\begin{table}[htbp]
\caption{
Expected signal ($S^\text{exp}_{68}$), background ($B^\text{exp}_{68}$) and observed event yields in data ($N_{68}$) in a window of width $w_{68}$, expected to contain 68\% of the signal. The expected signal and backgrounds are extracted from a signal-plus-background fit to the Asimov dataset (Section \ref{sec:result}). The signal uncertainty reflects the impact in $\mu$, while the background uncertainty accounts for the statistical uncertainty from the fit. The expected sensitivity is given by $S^\text{exp}_{68}/\sqrt{S^\text{exp}_{68} + B^\text{exp}_{68}}$.
The last row shows the inclusive results, calculated by summing the contributions from all categories. The inclusive sensitivity is derived by combining the category sensitivities in quadrature.}
\label{tab:SBtable}
\begin{center}
\begin{tabular}{lccrcc}
\hline
\hline
Category            & $S^\text{exp}_{68}$   & \makebox[3em][r]{}$B^\text{exp}_{68}$\makebox[2em][l]{} & $N_{68}$ &  $w_{68}$ [\GeV]   & $S^\text{exp}_{68}/\sqrt{S^\text{exp}_{68}+B^\text{exp}_{68}}$ \\
\hline
Lepton              & $1.5 \pm 1.1$   & \makebox[3em][r]{75.4} $\pm$ \makebox[2em][l]{2.8}    & 70      & 4.4 & 0.17   \\
VBFT                & $1.5 \pm 1.1$   & \makebox[3em][r]{1.3}  $\pm$ \makebox[2em][l]{0.4}    & 3       & 3.8 & 0.92   \\
VBFL                & $2.8 \pm 2.0$   & \makebox[3em][r]{27.2} $\pm$ \makebox[2em][l]{1.8}    & 18      & 4.0 & 0.52   \\
HRelpT-$ee$T        & $1.2 \pm 0.8$   & \makebox[3em][r]{6.6} $\pm$ \makebox[2em][l]{0.9}   & 9      & 3.2 & 0.43   \\
HRelpT-$ee$L        & $2.9 \pm 2.1$   & \makebox[3em][r]{53.9} $\pm$ \makebox[2em][l]{1.8}  & 61      & 4.0 & 0.39   \\
HRelpT-$\mu\mu$T    & $2.4 \pm 1.7$   & \makebox[3em][r]{20.3} $\pm$ \makebox[2em][l]{1.7}    & 25      & 3.9 & 0.51   \\
HRelpT-$\mu\mu$L    & $2.4 \pm 1.7$   & \makebox[3em][r]{56.5} $\pm$ \makebox[2em][l]{1.7}    & 61      & 4.1 & 0.32   \\
LRelpT-$ee$T        & $8.9 \pm 6.2$   & \makebox[3em][r]{231} $\pm$ \makebox[2em][l]{6}   & 240     & 3.8 & 0.57   \\
LRelpT-$ee$M        & $29 \pm 20$ & \makebox[3em][r]{2~562} $\pm$ \makebox[2em][l]{19}   & 2 587   & 4.1 & 0.56   \\
LRelpT-$ee$L        & $24 \pm 17$ & \makebox[3em][r]{13~122} $\pm$ \makebox[2em][l]{50}  & 13 074  & 4.5 & 0.21   \\
LRelpT-$\mu\mu$T    & $4.9 \pm 3.4$   & \makebox[3em][r]{95} $\pm$ \makebox[2em][l]{4}    & 107     & 3.9 & 0.49   \\
LRelpT-$\mu\mu$M    & $34 \pm 24$ & \makebox[3em][r]{2~527} $\pm$ \makebox[2em][l]{19}& 2 583   & 4.1 & 0.67   \\
LRelpT-$\mu\mu$L    & $36 \pm 25$ & \makebox[3em][r]{16~844} $\pm$ \makebox[2em][l]{40}   & 16 642  & 4.4 & 0.28   \\
\hline
Inclusive           & $150 \pm 110$ & \makebox[3em][r]{35~625} $\pm$ \makebox[2em][l]{70} & 35 480 & 4.0 & 1.81  \\
\hline
\hline
\end{tabular}
\end{center}
\end{table}
\end{DIFnomarkup}

%
%
%
%
%
%
%
%
%
%
%
%
%
%
%
%
%
%
%
%
%
%

%
%
%
%
%
%
%
%
%
%
%
%

%
%
%
%
%
%
%
%
%
%
%
%
%
%
%

% End of text imported from the .//sections/Categorization.tex input file

%
\FloatBarrier
\section{Signal and background modelling}
\label{sec:sig_bkg_modeling}

% The next lines are included from the .//sections/Signal_Modeling.tex input file
The signal and background yields are determined by an unbinned extended maximum-likelihood fit to the reconstructed \mzg spectrum in data, employing analytic functions for both components.
The shape of the $H\to Z\gamma$ signal is modelled by a double-sided Crystal Ball (DSCB) function, comprising a Gaussian core with power-law tails on both sides to capture both the detector resolution and non-Gaussian effects~\cite{dscb2}.
In each category, the DSCB parameters (mean $\mu_{\mathrm{CB}}$, width $\sigma_{\mathrm{CB}}$, and four parameters describing the tails) are determined from a fit to a combination of all signal samples.
The mean is shifted by 90~\MeV to correct for the generated Higgs boson mass of 125~\GeV to the measured value of 125.09~\GeV.
The overall signal normalisation and acceptance are taken directly from the same simulation.
A small contribution from $H\to\mu\mu$ decays (up to 3.8\% of \htoZg signal in certain categories) is likewise modelled with its own DSCB template, with normalisation fixed to the SM prediction. Depending on the event category, the fitted signal mass resolution ranges from 1.14 to 1.89 GeV.

% End of text imported from the .//sections/Signal_Modeling.tex input file

% The next lines are included from the .//sections/Background_Modeling.tex input file

In each category the background is modelled via an empirical analytical function that is chosen for its ability to provide a precise description of the total background. For all selected events, the backgrounds arise primarily from non-resonant \Zg production and \zjets events in which a jet is misidentified as a photon.
The diboson contribution remains below 0.2\% and is therefore neglected in the inclusive background estimate.
The relative contributions of the \Zg and \zjets backgrounds are obtained inclusively using a two-dimensional sideband (ABCD) method. Four regions are defined by (i) the photon identification requirement, \textit{tight} versus relaxed, and (ii) the isolation requirement, isolated versus non-isolated. The yield of the \Zg process in the signal region (\textit{tight} × isolated) is then extracted using the three control regions under the assumption that identification and isolation efficiencies factorise for both background components~\cite{STDM-2010-08}.
This yields a \Zg fraction of $0.49^{+0.05}_{-0.10}$, where the uncertainty is derived as the envelope of alternative estimations of the jet background correlations between regions.
Compared to $0.78_{-0.09}^{+0.04}$ in Run~2~\cite{HIGG-2018-42}, the reduction reflects three combined effects: the increased pile-up in Run~3, the lowered photon \pt threshold in the event selection, which admits more jets faking photons, and the smaller ratio of SM \Zg to \zjets cross-sections at \sqtts.

The background \mzg distribution is modelled based on fits to a dedicated background template (the underlying background distribution) that combines the \Zg and \zjets components.
After the inclusive fractions of \Zg and \zjets events are determined, they are extrapolated to each event category using efficiencies from the \Zg fast-simulation MC and data-driven \zjets samples.
To reduce statistical fluctuations in the control region, the ratio of \zjets to \Zg shapes as a function of \mzg is smoothed by fitting a low-order polynomial or exponential-polynomial function.
This smoothed ratio is applied to the \Zg template to define the \zjets background shape.
In the VBF categories, a similar smoothing is employed on the ratio of $\Zg+ (\zjets)$ to the larger EW \Zyjj simulation to define the total background template.
In the Lepton category, the 35\% contribution from diboson backgrounds, taken from simulation, is added to the template.
Finally, in each category, the overall background normalization is scaled to data in the \mzg sidebands (excluding the interval 120--130~\GeV). The resulting templates are compatible with the data \mzg distribution, yielding $p$-values ($\chi^2$) ranging from 1.2\% to 78\%.

To determine an analytical model that accurately represents the background template without inducing artificial signal features, a spurious signal ($SS$) study~\cite{HIGG-2018-42} is performed. In each category, several analytic function families, such as power laws, Bernstein polynomials, exponential polynomials, and logarithmic polynomials of the form $(1-x^{1/3})^fx^{\Sigma^{N}_{i=0} p_i\log(x)^i}$, with $x=m_{Z\gamma}/\sqrt{\mathrm{s}}$ and f a free parameter, are fitted to the background template.
Functions in each family are tested over three fit ranges (110--155~\GeV, 115--160~\GeV, 110--160~\GeV).
The function-range combinations are required to satisfy two criteria:
\begin{itemize}
\item The $|SS|$ is defined as the absolute value of the maximum fitted signal yield obtained from a background-only template when varying $m_H$ between 120 and 130~\GeV in steps of 1~\GeV. It is required that one of $|SS-1\sigma|$, $SS$, or $|SS+1\sigma|$ is less than $0.2\Delta S$,
where $\sigma$ is the error of the $SS$, and $\Delta S$ is the expected signal statistical uncertainty.
\item The $\chi^2$‐probability of the background-only fit $>1\%$.
\end{itemize}
When there is more than one function–range combination that passes these criteria, the one with the fewest free parameters is chosen. If more than one has the same number of parameters, the one with the smallest $|SS|$ is chosen.
If no candidate satisfies both criteria, the fit window is progressively narrowed (to a minimum width of 35~\GeV) until a valid model is found; if still unsuccessful, the combination with the lowest $|SS|$ is adopted.
A Wald test~\cite{Wald_test,HIGG-2020-16} is then applied to data sidebands to compare nested functions: if a lower-degree function is statistically compatible without significantly increasing $|SS|$, it replaces the nominal model to guard against over‐fitting.
To account for uncertainties in the relative \Zg and \zjets fractions, variations of those fractions by plus or minus their uncertainties are used to build alternative templates.
Repeating the $SS$ study on these templates leaves the chosen function–range combination unchanged in every category, and the maximum $|SS|$ across the three templates is assigned as the systematic uncertainty on the background model.
Once the function is selected, a smoothing technique based on Gaussian Process Regression~\cite{frate2017modelingsmoothbackgroundsgeneric, HIGG-2020-16} is applied to each category's template.
The $|SS|$ is then re-evaluated. This reduces the effect of residual statistical fluctuations without introducing any shape bias and does not affect the choice of any function–range combination in any categories. Unlike the signal models, whose parameters are fixed from simulation, the background model parameters are extracted directly from a fit to the \mzg distribution in data.
As an additional validation, a template is built as the sum of the SM signal template and the background template in each category, and is then fitted.
The resulting best‐fit signal strength in every category is compatible with the SM prediction, confirming the robustness of the background models.

%
%
%
%

% End of text imported from the .//sections/Background_Modeling.tex input file

%
\FloatBarrier
\section{Systematic uncertainties}
\label{sec:syst_uncert}

% The next lines are included from the .//sections/Systematic_Uncertainties.tex input file
Systematic uncertainties impact the fit results by affecting either the normalization or the shape of the $m_{Z\gamma}$ invariant mass distributions. Experimental uncertainties are evaluated using \RThr data and MC events.
Despite the large simulation background samples and smoothing procedures, the spurious signal, uncorrelated across event categories, retains a large contribution from statistical fluctuations of the input samples. It results in an 11\% impact on the signal strength $\mu$, which is small with respect to the statistical uncertainty. %

Signal shape modelling uncertainties vary by category and are driven by the calibration of the electron and photon energy, as well as the muon momentum.
The uncertainty in the electron and photon energy resolution and in the muon momentum resolution leads to a less than 5\% mass resolution ($\sigma_{CB}$) uncertainty. The uncertainty in the electron and photon energy scale (and in the muon momentum scale) leads to a less than 0.3\% (0.1\%) uncertainty in the peak position ($\mu_{CB}$).
Overall, these uncertainties affect the measured signal strength by less than 2\%.

The 60\% uncertainty in the $H\rightarrow\mu\mu$ contribution, taken from the Run-2 ATLAS measurement~\cite{HIGG-2019-14}, contributes a 2\% uncertainty in the expected signal strength.
Additional uncertainties in the signal yield arising from reconstruction, identification, and isolation corrections are smaller than 3\% for photons, electrons, and muons~\cite{EGAM-2021-01,EGAM-2021-02}.
The jet-related uncertainties, including those from jet reconstruction and in situ calibration, are each below 2\%.
A conservative uncertainty of 10\% is assigned to the reconstruction and calibration of forward jets.
All of the jet uncertainties translate into a 4\% impact on the signal strength.
The uncertainty related to the \pileup effects is negligible.
The overall uncertainty in the integrated luminosity for each dataset using the LUCID-2 detector~\cite{LUCID2} is 4\%\footnote{The integrated luminosity for the 2024 dataset is derived from the relative yields of $Z \to ee$ and $Z \to \mu\mu$ events relative to those measured in 2022/2023, yielding a preliminary uncertainty of 5\%.
When combined with the 2\% uncertainty assigned to the 2022/2023 integrated luminosity~\cite{ATL-DAPR-PUB-2023-001,ATL-DAPR-PUB-2024-001}, a luminosity-weighted uncertainty of 4\% is obtained for the \RThr dataset.} for 2022--2024.
Additional uncertainties arising from the triggers are below 2\% across the analysis categories.
Combining all experimental systematics, excluding the spurious signal uncertainty, yields a total 9\% uncertainty in the signal strength.

Theory uncertainties affecting the expected signal yield arise from multiple sources.
Uncertainties in the production cross-section and kinematic distributions, primarily due to missing higher-order QCD corrections, contribute up to 12\%, depending on the analysis category, and are dominated by variations in the renormalisation and factorisation scales.
The parton shower modelling uncertainty, evaluated by comparing \PYTHIAV{8} and \HERWIGV{7}, ranges from 3\% to 29\%, with the largest effects observed in the VBF categories. In the non-VBF categories, the corresponding uncertainty is typically between 0.3\% and 10\%.
The uncertainty in the Higgs boson branching ratio to $Z\gamma$ is 7\%~\cite{deFlorian:2016spz}. The total impact of theory uncertainties on the signal strength is estimated to be 12\%.
Table~\ref{tab:uncertainty_breakdown_run3} summarises the symmetrised impacts of individual uncertainty sources on the expected and observed signal strengths. The total uncertainty is dominated by the statistical component, with an expected impact of 0.68. Systematic uncertainties contribute 0.16, the spurious signal and theory modelling effects being the leading sources.

\begin{table}[htbp]
\centering
\caption{Breakdown of the symmetrised impacts from individual sources of uncertainty on the expected and observed signal strengths. Uncertainties are grouped by source type to illustrate their relative contributions.}
\begin{tabular}{lcc}
\hline
\multirow{2}{*}{Uncertainty source} & \multicolumn{2}{c}{$\Delta\mu$} \\
\cline{2-3}                       &  Expected & Observed\\
\hline
Statistical uncertainty           & $0.68$ & $0.65$ \\
Systematic uncertainty            & $0.16$ & $0.17$ \\
\quad Spurious signal (background modelling)             & $0.11$ & $0.11$ \\
\quad QCD scale, PDF+$\alpha_{S}$, parton shower   & $0.09$ & $0.06$ \\
\quad Branching ratio (\htoZg)             & $0.05$ & $0.05$ \\
\quad Luminosity                  & $0.03$ & $0.03$ \\
\quad Photon efficiency           & $0.02$ & $0.03$ \\
\quad Jet                         & $0.02$ & $0.07$ \\
\quad Electron and photon energy scale and resolution        & $0.02$ & $0.06$ \\
\quad Electron efficiency         & $0.02$ & $0.02$ \\
\quad Muon                        & $<0.01$ & $0.02$ \\
\quad Trigger                     & $<0.01$ & $0.02$ \\
\hline
Total                             & $0.70$ & $0.67$ \\
\hline
\end{tabular}
\label{tab:uncertainty_breakdown_run3}
\end{table}

%

%
%
%
%
%
%
%
%
%
%
%
%
%
%
%
%
%
%
%

% End of text imported from the .//sections/Systematic_Uncertainties.tex input file

%
\FloatBarrier
\section{Results}
\label{sec:result}

% The next lines are included from the .//sections/Results.tex input file
\subsection{\RThr only}

The results are extracted via an unbinned, simultaneous maximum-likelihood fit~\cite{Cowan:2010js} to the \mzg distributions across all categories, each with its own optimised mass window determined as described in Section~\ref{sec:sig_bkg_modeling}, following the methodology of previous \HZy searches~\cite{HIGG-2013-05, HIGG-2018-42}.
The likelihood function is built by incorporating both the signal strength of \htoZg and the nuisance parameters (NPs), which characterise the effects of systematic uncertainties on the signal normalisation and shape, as well as background normalisation and shape parameters.

The measured signal strength is $\mu = 0.9 ^{+0.7}_{-0.6}(\text{stat.})^{+0.2}_{-0.1}(\text{syst.}) = 0.9_{-0.6}^{+0.7}$ for $m_H=125.09$~\GeV.
Under the SM signal hypothesis, the expected signal strength is $\mu_\mathrm{exp} = 1.0 \pm 0.7(\text{stat.})^{+0.2}_{-0.1}(\text{syst.})=1.0\pm0.7$.
The corresponding Asimov dataset~\cite{Cowan:2010js} is generated under the hypothesis of SM signal plus background.
The result is statistically compatible with the SM expectation.
The observed significance under the background-only hypothesis is 1.4$\sigma$, close to the expected significance of 1.5$\sigma$. Among the event categories, the VBFT category provides the highest expected sensitivity. In all cases, the uncertainties are dominated by the statistical contribution, while the most significant systematic uncertainty impact arises from the modelling of the background (spurious signal).
The measurement using individual signal strengths for each category is consistent with the result obtained from the global signal strength, with a $p$-value of 0.35.

\begin{figure}[htbp]
\centering
\begin{minipage}[b]{0.48\textwidth}
\centering
\includegraphics[width=\textwidth]{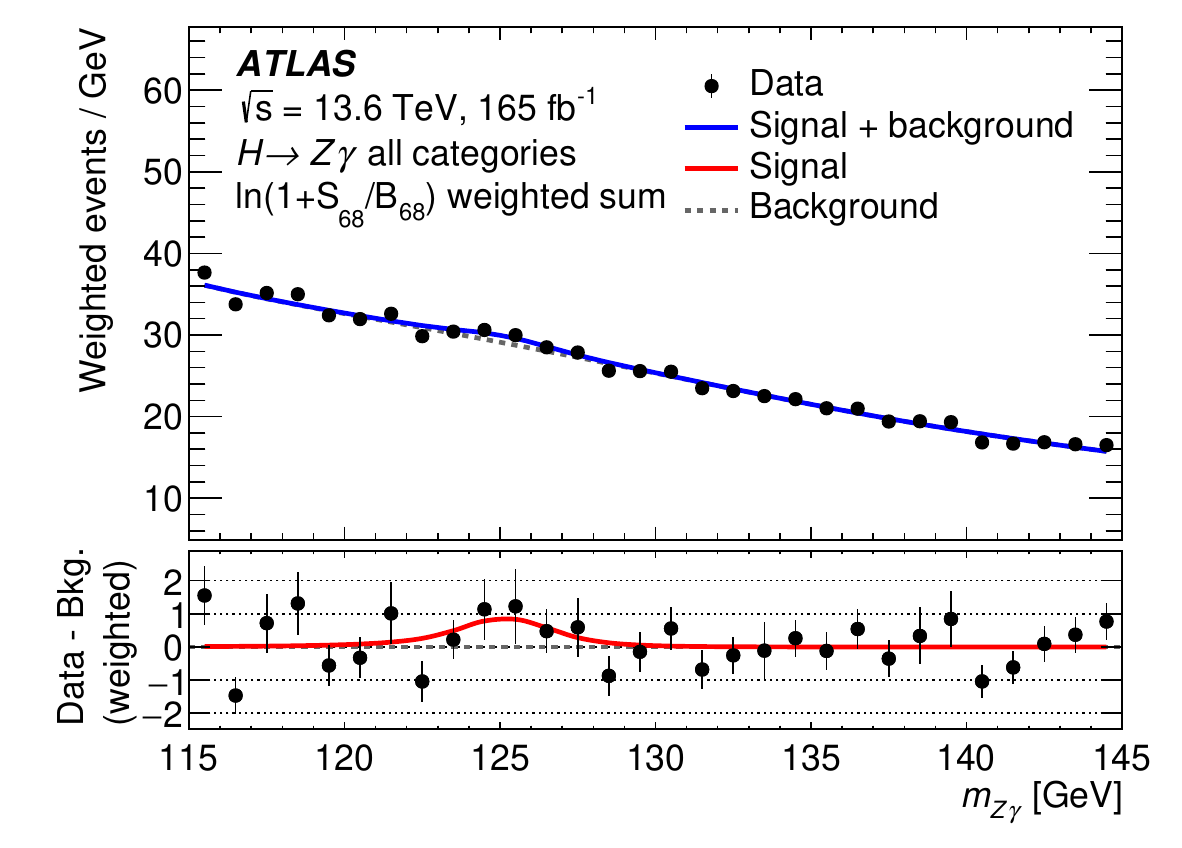}
\caption*{(a)}
\end{minipage}
\hfill
\begin{minipage}[b]{0.48\textwidth}
\centering
\includegraphics[width=\textwidth]{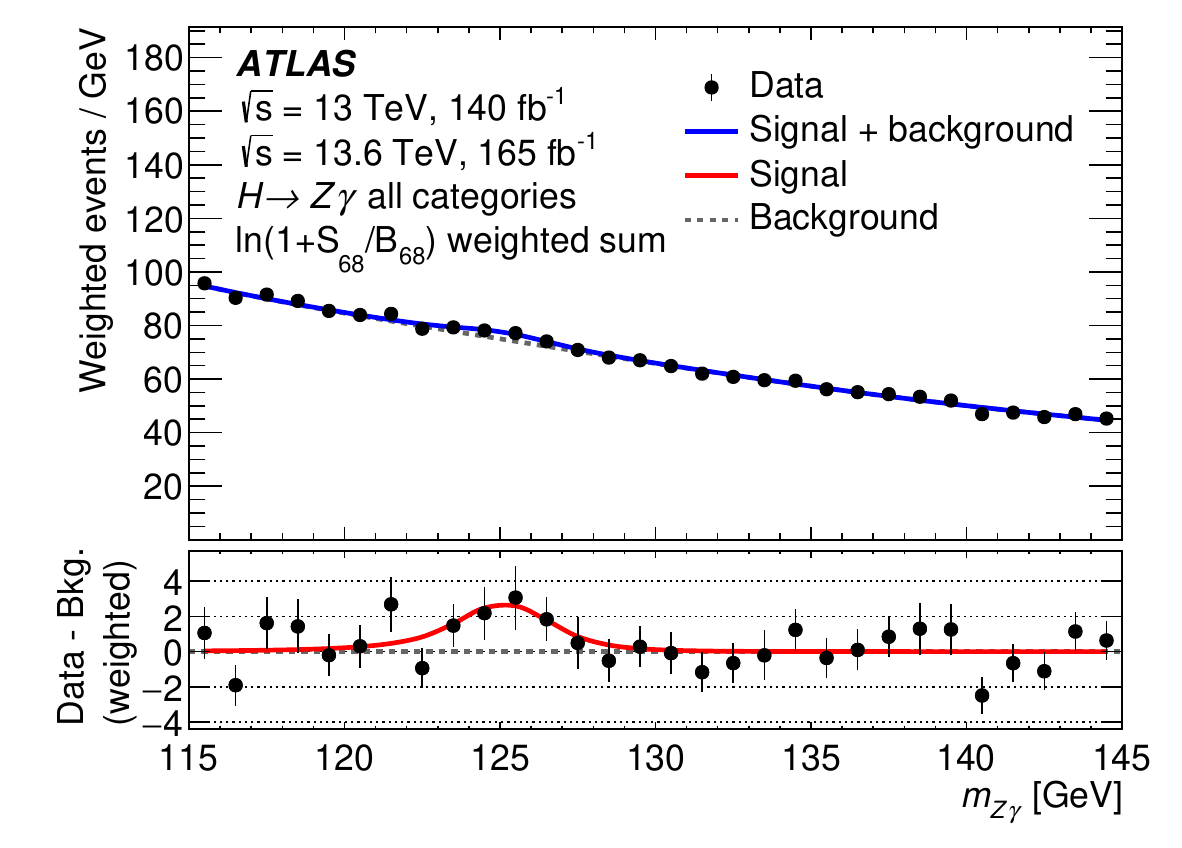}
\caption*{(b)}
\end{minipage}
\caption{\Zg invariant mass distributions of weighted data events across all categories for (a) \RThr only and (b) the combined \RTwo and \RThr dataset.
The black points represent the data, with statistical uncertainties shown as error bars.
Each event is weighted by $\ln(1+S_{68}/B_{68})$, where $S_{68}$ and $B_{68}$ are the signal and background yields in each category, estimated from the fit to the data, in an \mzg window expected to contain 68\% of the signal.
The signal-plus-background fit (solid blue curve) and the background model (dashed line) are overlaid. In the bottom panels, the residuals between the data and the background model (black dots with error bars) are compared to the signal model (red solid line).
All curves represent the weighted sum of the individual category models.}
\label{fig:mllg-soverb}
\end{figure}

\begin{figure}[htbp]
\centering
\begin{tabular}{cc}
\includegraphics[width=0.48\textwidth]{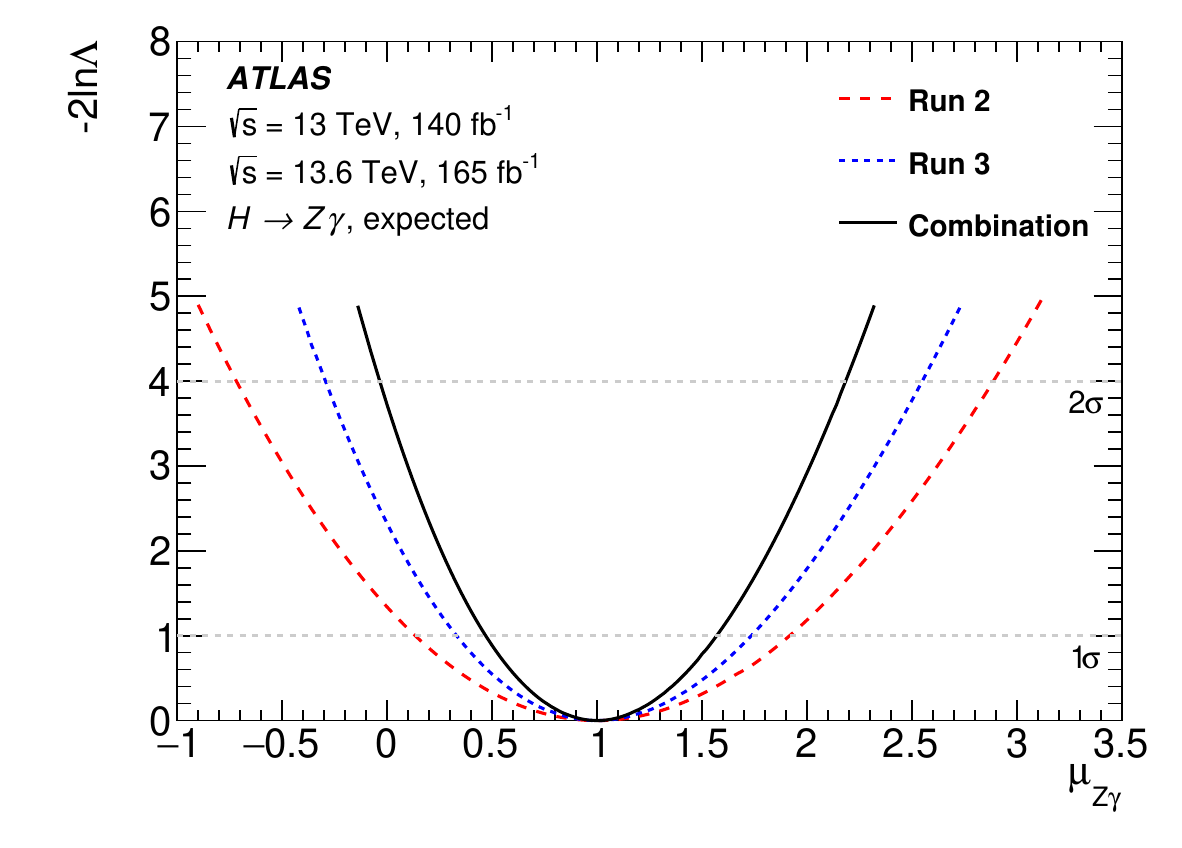} &
\includegraphics[width=0.48\textwidth]{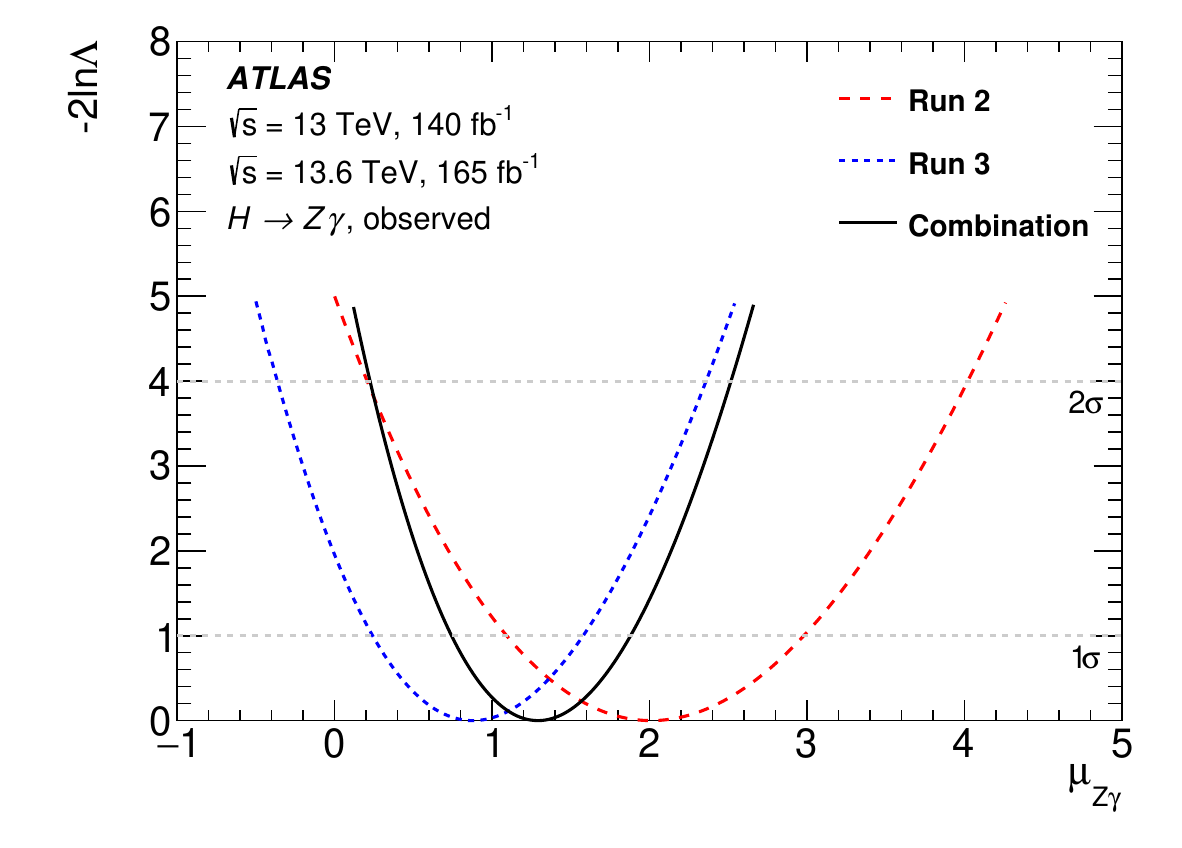} \\
(a)& (b)
\end{tabular}
\caption{Profiled likelihood scan of the signal strength for the (a) expected and (b) observed results. Both plots show the results obtained by this analysis of the \RThr dataset (dashed blue line), by the previous analysis of the \RTwo dataset~\cite{HIGG-2018-42} (long dashed red line), and their combination (solid black line).}
\label{fig:likelihood-scan}
\end{figure}

Figure~\ref{fig:mllg-soverb} (a) presents the \mzg distribution of selected events in the \RThr dataset, weighted by $\ln(1+S_{68}/B_{68})$, where $S_{68}$ and $B_{68}$ are the signal and background yields in each category, estimated from the fit to the data, in an \mzg window expected to contain 68\% of the signal.
The combined models, obtained by weighting the individual category models, are overlaid.
The negative profiled-likelihood-ratio values as a function of the signal strength $\mu$ are shown in Figure~\ref{fig:likelihood-scan}.

Overall, this analysis achieves an expected significance improvement of 27\% relative to the previous ATLAS result~\cite{HIGG-2018-42}.
This gain is primarily driven by the advanced event selection and categorisation, which contributes 14\%. The remaining improvement is from the larger dataset and a more favourable signal-to-background ratio at $\sqrt{s}=13.6$~\TeV.

\subsection{\RTwo and \RThr combination}

These results based on \RThr data are combined with those from the previous ATLAS search for $H\to Z\gamma$ decays using \RTwo data~\cite{HIGG-2018-42}.
The theoretical uncertainties, including branching ratios, QCD scale, and $\alpha_s$ uncertainties, are treated as fully correlated across both datasets.
The PDF uncertainties are uncorrelated due to the distinct PDF sets used in Run~2 and Run~3.
Due to the different data-taking conditions, most experimental uncertainties are treated as uncorrelated between the two runs, in particular, the uncertainties related to the trigger, luminosity, electron, photon, and spurious signal. All of the uncertainty sources from muons and jets are correlated, except for the uncorrelated in situ calibration uncertainties.
The impact of the correlation scheme of systematic uncertainties on the final results was studied and found to be negligible, as the overall uncertainty is dominated by the statistical contributions.
The joint likelihood is constructed as the product of the individual likelihood functions from the \RTwo and \RThr analyses. Correlated systematic sources share a common NP with a Gaussian constraint in the combined fit.

\begin{figure}[htbp]
\centering
\begin{subfigure}[t]{0.80\textwidth}
\centering
\includegraphics[width=\linewidth]{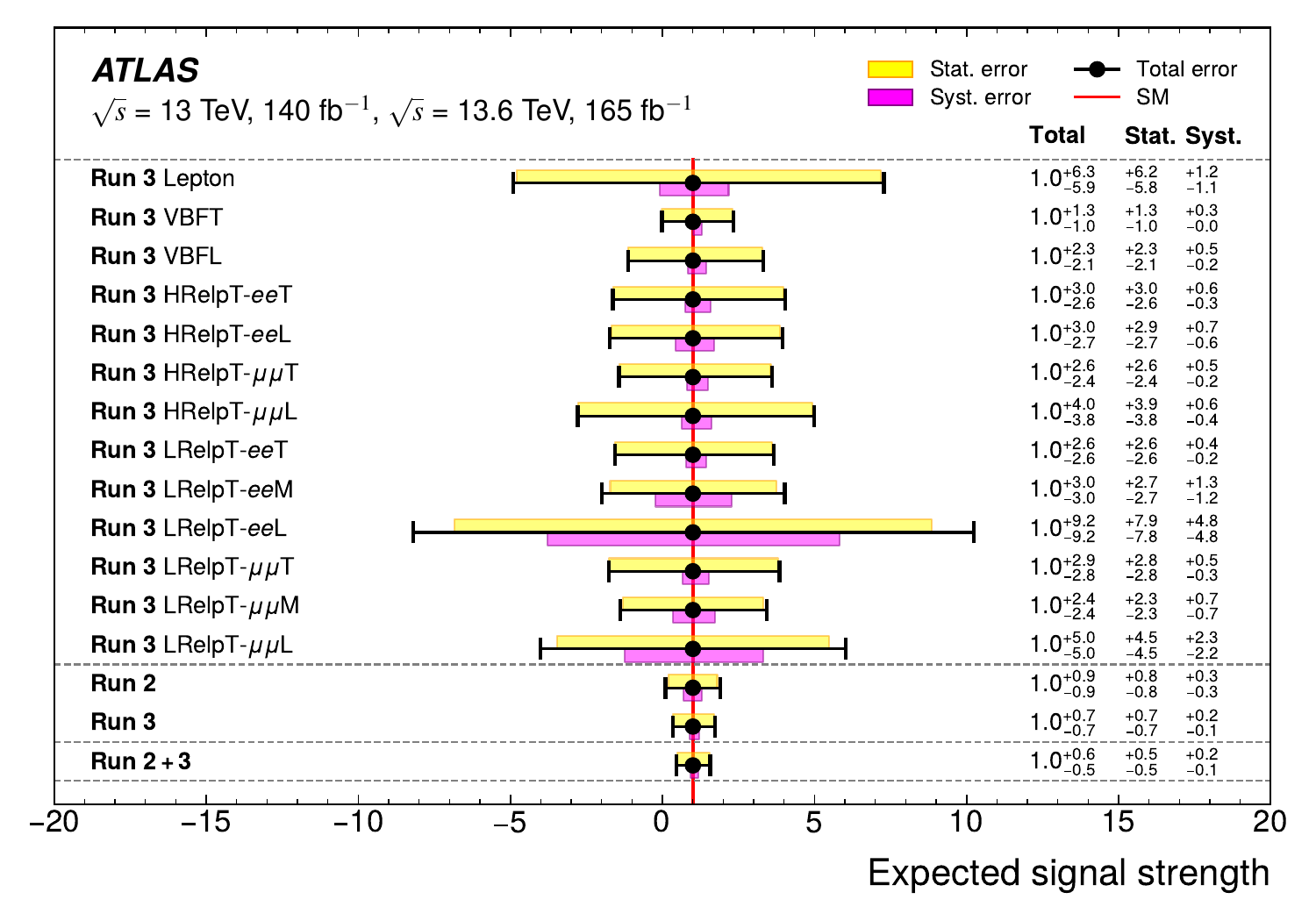}
\caption{}
\label{fig:mu_summary_exp}
\end{subfigure}
\\
\begin{subfigure}[t]{0.80\textwidth}
\centering
\includegraphics[width=\linewidth]{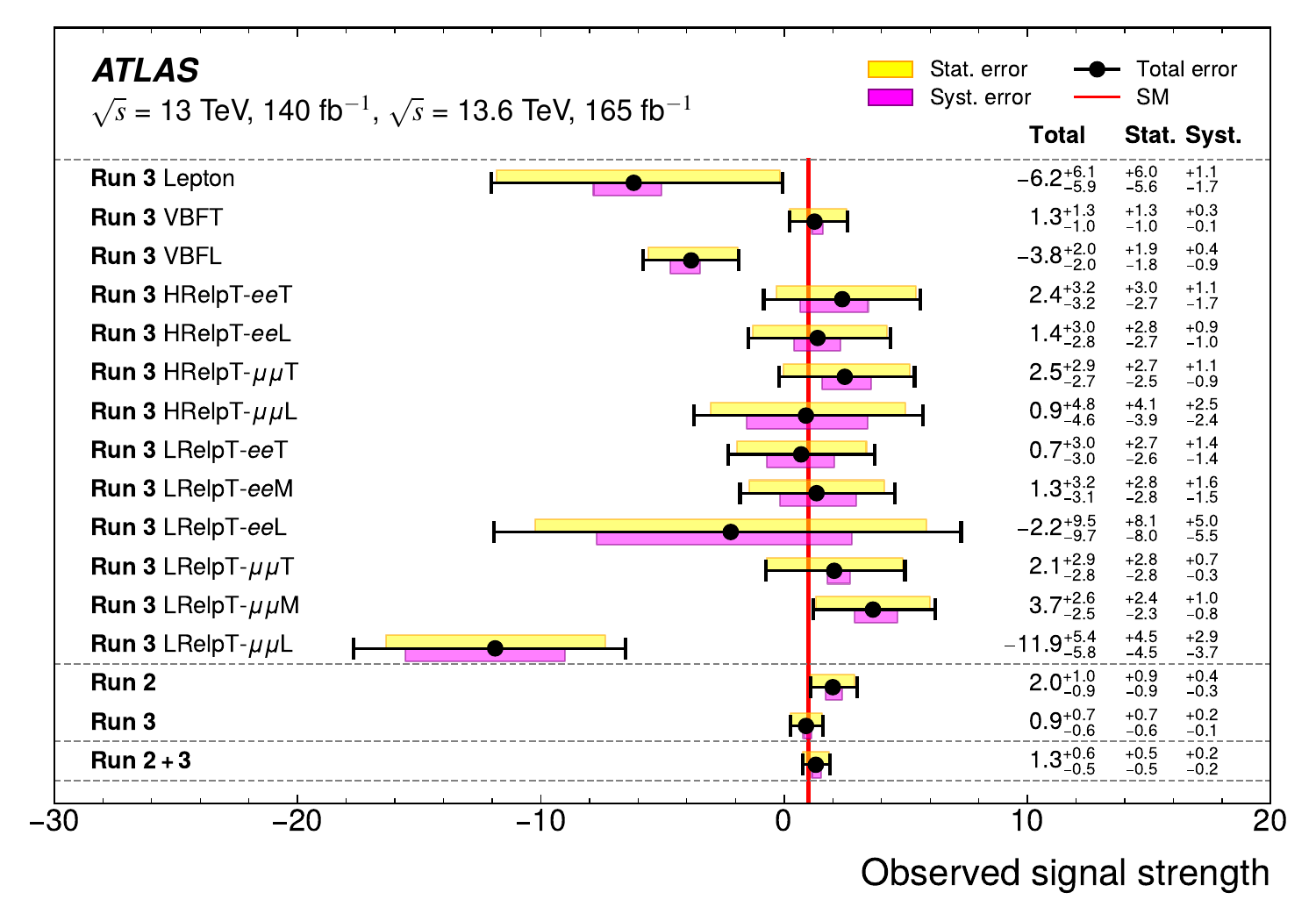}
\caption{}
\label{fig:mu_summary_obs}
\end{subfigure}
\caption{(a) Expected and (b) observed best-fit signal strengths and corresponding uncertainties in each category of the \RThr analysis, their combined values, the results of the previous \RTwo analysis, and the overall combination. The black points with horizontal error bars correspond to the central values and total uncertainties, while the upper yellow (lower magenta) bands indicate the statistical (systematic) uncertainties. The red vertical line denotes the SM prediction.}
\label{fig:NLL_aux}
\end{figure}

Figure~\ref{fig:mllg-soverb} (b) presents the weighted \mzg distribution with the fitted signal and background models overlaid, and Figure~\ref{fig:likelihood-scan} shows the likelihood scan of the combined signal strength. The observed combined fit has a best-fit signal strength of $\mu=1.3\pm0.5 (\mathrm{stat.})\pm0.2(\mathrm{syst.})=1.3^{+0.6}_{-0.5}$ with an expected $\mu=1.0\pm0.5 (\mathrm{stat.})^{+0.2}_{-0.1}(\mathrm{syst.})=1.0^{+0.6}_{-0.5}$.
The significance of the observed (expected) excess above the background-only hypothesis is 2.5$\sigma$ (1.9$\sigma$). Relative to the \RTwo analysis, the combined expected significance improves by 61\%.
Figure \ref{fig:NLL_aux} shows the signal strength results in each \RThr category, their combined value, the results of the previous Run-2 analysis, and the overall combination.
The \RTwo and \RThr measurements are compatible,  with a $p$-value of 0.32.
Assuming the SM production cross-sections,
the combination provides the most stringent expected sensitivity to date for determining the \htoZg branching fraction, surpassing the ATLAS + CMS \RTwo combination~\cite{HIGG-2022-22-fixed}.
Under this assumption, the observed (expected) branching fraction is $(2.0_{-0.8}^{+0.9})\times 10^{-3}$ ($(1.5_{-0.8}^{+0.9})\times 10^{-3}$), compared with the SM prediction of $(1.54^{+0.10}_{-0.11})\times10^{-3}$.

% End of text imported from the .//sections/Results.tex input file

%
\FloatBarrier

\section{Conclusion}
\label{sec:conclusion}

% The next lines are included from the .//sections/Conclusion.tex input file

A search for the Higgs boson decay to \Zg in the \lly final state is performed using proton--proton collision data recorded with the ATLAS detector at $\sqrt{s}=13.6$~\TeV during 2022--2024, corresponding to an integrated luminosity of 165~\invfb.
A simultaneous unbinned maximum-likelihood fit to the reconstructed invariant mass of the \Zg system across all event categories gives an observed (expected) signal yield normalised to the SM prediction, $\mu=0.9^{+0.7}_{-0.6}$ ($\mu=1.0\pm0.7$). This corresponds to an observed (expected) signal significance of 1.4 (1.5) standard deviations. The statistical uncertainty is the dominant source of error.
Compared to a similar search performed with \RTwo data~\cite{HIGG-2018-42}, the expected significance improves by 28\%,
with 15\% arising from the enhanced event selection and categorisation strategies, and the remainder from increased integrated luminosity and larger cross-sections.

This measurement is further combined with the \RTwo result to yield the most stringent expected sensitivity to date for the \htoZg decay, surpassing the ATLAS + CMS \RTwo combination~\cite{HIGG-2022-22-fixed}.
In the combined fit, the observed (expected) signal yield normalised to the SM prediction is $\mu=1.3^{+0.6}_{-0.5}$ ($\mu=1.0^{+0.6}_{-0.5}$), corresponding to an observed (expected) signal significance of 2.5 (1.9) standard deviations.
The combination improves the expected significance by 61\% with respect to the \RTwo measurement.
The measurement is consistent with the SM expectation, indicating no significant deviation from the predicted Higgs boson behaviour in the \Zg channel.

% End of text imported from the .//sections/Conclusion.tex input file

%
%
%
%
%
%
\section*{Acknowledgements}
%

%

% The next lines are included from the .//acknowledgements/Acknowledgements.tex input file
%
%

%
%

We thank CERN for the very successful operation of the LHC and its injectors, as well as the support staff at
CERN and at our institutions worldwide without whom ATLAS could not be operated efficiently.

The crucial computing support from all WLCG partners is acknowledged gratefully, in particular from CERN, the ATLAS Tier-1 facilities at TRIUMF/SFU (Canada), NDGF (Denmark, Norway, Sweden), CC-IN2P3 (France), KIT/GridKA (Germany), INFN-CNAF (Italy), NL-T1 (Netherlands), PIC (Spain), RAL (UK) and BNL (USA), the Tier-2 facilities worldwide and large non-WLCG resource providers. Major contributors of computing resources are listed in Ref.~\cite{ATL-SOFT-PUB-2025-001}.

We gratefully acknowledge the support of ANPCyT, Argentina; YerPhI, Armenia; ARC, Australia; BMWFW and FWF, Austria; ANAS, Azerbaijan; CNPq and FAPESP, Brazil; NSERC, NRC and CFI, Canada; CERN; ANID, Chile; CAS, MOST and NSFC, China; Minciencias, Colombia; MEYS CR, Czech Republic; DNRF and DNSRC, Denmark; IN2P3-CNRS and CEA-DRF/IRFU, France; SRNSFG, Georgia; BMFTR, HGF and MPG, Germany; GSRI, Greece; RGC and Hong Kong SAR, China; ICHEP and Academy of Sciences and Humanities, Israel; INFN, Italy; MEXT and JSPS, Japan; CNRST, Morocco; NWO, Netherlands; RCN, Norway; MNiSW, Poland; FCT, Portugal; MNE/IFA, Romania; MSTDI, Serbia; MSSR, Slovakia; ARIS and MVZI, Slovenia; DSI/NRF, South Africa; MICIU/AEI, Spain; SRC and Wallenberg Foundation, Sweden; SERI, SNSF and Cantons of Bern and Geneva, Switzerland; NSTC, Taipei; TENMAK, T\"urkiye; STFC/UKRI, United Kingdom; DOE and NSF, United States of America.

Individual groups and members have received support from BCKDF, CANARIE, CRC and DRAC, Canada; CERN-CZ, FORTE and PRIMUS, Czech Republic; COST, ERC, ERDF, Horizon 2020, ICSC-NextGenerationEU and Marie Sk{\l}odowska-Curie Actions, European Union; Investissements d'Avenir Labex, Investissements d'Avenir Idex and ANR, France; DFG and AvH Foundation, Germany; Herakleitos, Thales and Aristeia programmes co-financed by EU-ESF and the Greek NSRF, Greece; BSF-NSF and MINERVA, Israel; NCN and NAWA, Poland; La Caixa Banking Foundation, CERCA Programme Generalitat de Catalunya and PROMETEO and GenT Programmes Generalitat Valenciana, Spain; G\"{o}ran Gustafssons Stiftelse, Sweden; The Royal Society and Leverhulme Trust, United Kingdom.

In addition, individual members wish to acknowledge support from Australia: Funding agency (AGAUR - 2023 BP 00141); CERN: European Organization for Nuclear Research (CERN DOCT); Chile: Agencia Nacional de Investigaci\'on y Desarrollo (FONDECYT 1230812, FONDECYT 1240864, Fondecyt 3240661, Fondecyt Regular 1240721); China: Chinese Ministry of Science and Technology (MOST-2023YFA1605700, MOST-2023YFA1609300), National Natural Science Foundation of China (NSFC - 12175119, NSFC 12275265); Czech Republic: Czech Science Foundation (GACR - 24-11373S), Ministry of Education Youth and Sports (ERC-CZ-LL2327, FORTE CZ.02.01.01/00/22\_008/0004632), PRIMUS Research Programme (PRIMUS/21/SCI/017); EU: H2020 European Research Council (ERC - 101002463); European Union: European Research Council (BARD No. 101116429, ERC - 948254, ERC 101089007), European Regional Development Fund (HE COFUND GA No.101081355, ERDF), Horizon 2020 Framework Programme (MUCCA - CHIST-ERA-19-XAI-00), European Union, Future Artificial Intelligence Research (FAIR-NextGenerationEU PE00000013), Italian Center for High Performance Computing, Big Data and Quantum Computing (ICSC, NextGenerationEU); France: Agence Nationale de la Recherche (ANR-21-CE31-0013, ANR-21-CE31-0022, ANR-22-EDIR-0002, ANR-24-CE31-0504-01); Germany: Baden-Württemberg Stiftung (BW Stiftung-Postdoc Eliteprogramme), Deutsche Forschungsgemeinschaft (DFG - 469666862, DFG - CR 312/5-2); China: Research Grants Council (GRF); Italy: Istituto Nazionale di Fisica Nucleare (ICSC, NextGenerationEU), Ministero dell'Università e della Ricerca (NextGenEU 153D23001490006 M4C2.1.1, NextGenEU I53D23000820006 M4C2.1.1, NextGenEU I53D23001490006 M4C2.1.1, SOE2024\_0000023); Japan: Japan Society for the Promotion of Science (JSPS KAKENHI JP22H01227, JSPS KAKENHI JP22H04944, JSPS KAKENHI JP22KK0227, JSPS KAKENHI JP24K23939, JSPS KAKENHI JP24KK0251, JSPS KAKENHI JP25H00650, JSPS KAKENHI JP25H01291, JSPS KAKENHI JP25K01023); Norway: Research Council of Norway (RCN-314472); Poland: Ministry of Science and Higher Education (IDUB AGH, POB8, D4 no 9722), Polish National Science Centre (NCN 2021/42/E/ST2/00350, NCN OPUS 2023/51/B/ST2/02507, NCN OPUS nr 2022/47/B/ST2/03059, NCN UMO-2019/34/E/ST2/00393, UMO-2022/47/O/ST2/00148, UMO-2023/49/B/ST2/04085, UMO-2023/51/B/ST2/00920, UMO-2024/53/N/ST2/00869); Portugal: Foundation for Science and Technology (FCT); Spain: Generalitat Valenciana (ASFAE/2022/008), Ministry of Science and Innovation (MCIN \& NextGenEU PCI2022-135018-2, MICIN \& FEDER PID2021-125273NB, RYC2019-028510-I, RYC2020-030254-I, RYC2021-031273-I, RYC2022-038164-I), Ministerio de Ciencia, Innovación y Universidades/Agencia Estatal de Investigaci\'on (PID2022-142604OB-C22); Sweden: Carl Trygger Foundation (Carl Trygger Foundation CTS 22:2312), Swedish Research Council (Swedish Research Council 2023-04654, VR 2021-03651, VR 2022-03845, VR 2022-04683, VR 2023-03403, VR 2024-05451), Knut and Alice Wallenberg Foundation (KAW 2018.0458, KAW 2022.0358, KAW 2023.0366); Switzerland: Swiss National Science Foundation (SNSF - PCEFP2\_194658); United Kingdom: Royal Society (NIF-R1-231091); United States of America: U.S. Department of Energy (ECA DE-AC02-76SF00515), Neubauer Family Foundation.

%
%

% End of text imported from the .//acknowledgements/Acknowledgements.tex input file

%
\clearpage
\printbibliography
\clearpage
% ATLAS Collaboration author list
% Reference date of HIGP-2024-19 is 2025-06-18
% Author list last updated on date 27-JAN-26
% Data extracted on 27-Jan-2026 for paper reference HIGP-2024-19
% at 10:35am
 
\begin{flushleft}
\hypersetup{urlcolor=black}
{\Large The ATLAS Collaboration}

\bigskip

\AtlasOrcid[0000-0002-6665-4934]{G.~Aad}$^\textrm{\scriptsize 104}$,
\AtlasOrcid[0000-0001-7616-1554]{E.~Aakvaag}$^\textrm{\scriptsize 17}$,
\AtlasOrcid[0000-0002-5888-2734]{B.~Abbott}$^\textrm{\scriptsize 123}$,
\AtlasOrcid[0000-0002-0287-5869]{S.~Abdelhameed}$^\textrm{\scriptsize 119a}$,
\AtlasOrcid[0000-0002-1002-1652]{K.~Abeling}$^\textrm{\scriptsize 55}$,
\AtlasOrcid[0000-0001-5763-2760]{N.J.~Abicht}$^\textrm{\scriptsize 49}$,
\AtlasOrcid[0000-0002-8496-9294]{S.H.~Abidi}$^\textrm{\scriptsize 30}$,
\AtlasOrcid[0009-0003-6578-220X]{M.~Aboelela}$^\textrm{\scriptsize 45}$,
\AtlasOrcid[0000-0002-9987-2292]{A.~Aboulhorma}$^\textrm{\scriptsize 36e}$,
\AtlasOrcid[0000-0001-5329-6640]{H.~Abramowicz}$^\textrm{\scriptsize 157}$
% ...
\bigskip
\\

% ...
$^{17}$Department for Physics and Technology, University of Bergen, Bergen; Norway.\\
% ...
$^{30}$Physics Department, Brookhaven National Laboratory, Upton NY; United States of America.\\
% ...
$^{36}$$^{(a)}$Facult\'e des Sciences Ain Chock, Universit\'e Hassan II de Casablanca;$^{(b)}$Facult\'{e} des Sciences, Universit\'{e} Ibn-Tofail, K\'{e}nitra;$^{(c)}$Facult\'e des Sciences Semlalia, Universit\'e Cadi Ayyad, LPHEA-Marrakech;$^{(d)}$LPMR, Facult\'e des Sciences, Universit\'e Mohamed Premier, Oujda;$^{(e)}$Facult\'e des sciences, Universit\'e Mohammed V, Rabat;$^{(f)}$Institute of Applied Physics, Mohammed VI Polytechnic University, Ben Guerir; Morocco.\\
% ...
$^{45}$Physics Department, Southern Methodist University, Dallas TX; United States of America.\\
% ...
$^{49}$Fakult\"{a}t Physik, Technische Universit{\"a}t Dortmund, Dortmund; Germany.\\
% ...
$^{55}$II. Physikalisches Institut, Georg-August-Universit\"{a}t G\"ottingen, G\"ottingen; Germany.\\
% ...
$^{104}$CPPM, Aix-Marseille Universit\'e, CNRS/IN2P3, Marseille; France.\\
% ...
$^{119}$$^{(a)}$New York University Abu Dhabi, Abu Dhabi;$^{(b)}$United Arab Emirates University, Al Ain; United Arab Emirates.\\
% ...
$^{123}$Homer L. Dodge Department of Physics and Astronomy, University of Oklahoma, Norman OK; United States of America.\\
% ...
$^{157}$Raymond and Beverly Sackler School of Physics and Astronomy, Tel Aviv University, Tel Aviv; Israel.\\
% ...
$^{a}$ Also at Affiliated with an institute formerly covered by a cooperation agreement with CERN.\\
$^{b}$ Also at An-Najah National University, Nablus; Palestine.\\
$^{c}$ Also at Borough of Manhattan Community College, City University of New York, New York NY; United States of America.\\
$^{d}$ Also at Center for Interdisciplinary Research and Innovation (CIRI-AUTH), Thessaloniki; Greece.\\
$^{e}$ Also at Centre of Physics of the Universities of Minho and Porto (CF-UM-UP); Portugal.\\
$^{f}$ Also at CERN, Geneva; Switzerland.\\
% ...

\end{flushleft}

% Created with Glance <Atlas.Glance@cern.ch>

%

%
%

%

%
%
%
%

\end{document}